\documentclass[letter,fleqn,11pt]{article}

\usepackage{graphicx}

\begin{document}
\noindent
\begin{center}
{\Large\bf Flow Patterns of Cellular Automata and Optimal-velocity
Traffic Models at Highway Bottlenecks}
\end{center}

\vspace{0.5cm}

\begin{center}
Peter {\sc Berg}$^1$ and Justin {\sc Findlay}

\vspace{0.5cm}

Faculty of Science, University of Ontario
Institute of Technology, \\
2000 Simcoe Street N., Oshawa, ON, L1H 7K4, Canada, \\
$^1$Email: peter.berg@uoit.ca, Phone: +1 905 721 8668,\\
FAX: +1 905 721 3304, web: www.peterberg.net

\vspace{0.5cm}

16 July 2007

\vspace{1.5cm}

{\bf Abstract}
\end{center}
A bottleneck simulation of road traffic on a loop, using the deterministic
cellular automata (CA) Nagel-Schreckenberg model with zero dawdling
probability, reveals three types of stationary wave solutions. They consist
of i) two shock waves, one at each bottleneck boundary, ii) one shock wave
at the boundary and one on the ``open'' road, and iii) the trivial solution,
i.e.~homogeneous, uniform flow. These solutions are selected dynamically from
a range of kinematicly permissible stationary shocks. This is
similar in fashion to the wave selection in a bottleneck simulation of the
optimal-velocity (OV) model, which is explained by a travelling wave
phase-plane analysis of the corresponding continuum model. It is yet another
strong indication that CA and OV models share certain underlying dynamics,
although the former are discrete in space and time while the latter are
continuous.

\vspace{0.5cm}

\cleardoublepage

\section{Introduction}
\label{intro}
Cellular automata (CA) models have been widely used to simulate traffic flow on
highways and road networks
\cite{schad:00,kerner:02a,yukawa:94,popkov:99,kolo:98,chey:01,chey:01b,barlo:02,
lakatos:05}, in particular the Nagel-Schreckenberg model \cite{nagel:92,schad:93}.
Together with car-following (CF) and continuum models, they represent the three
predominant classes of traffic models.

Analytical work by Berg {\em et al.} \cite{berg:00} and Lee {\em et al.}
\cite{lee:01} has established a link between car-following models based on
ordinary differential equations, and continuum models based on partial
differential equations. While an analytical link between CA models and either
CF or continuum models is still missing (mean-field theory aside), the
dynamics of all three classes exhibit many common features such as
sub-critical bifurcations, limit cycles and pattern formation
\cite{helbing:book,kerner:book}.

Bottlenecks are the major cause for highway congestion and, therefore, have
been studied in some detail
\cite{kerner:02b,jia:04,sugiyama:05,kerner:03,berg:01,helbing:00}.
In this paper, a wave selection analysis of a bottleneck simulation reveals a
fundamental link between the dynamics of CA models and optimal-velocity (OV)
models \cite{bando:95}, which belong to the class of CF models.

The paper is organised as follows.
We begin with the description of the CA bottleneck simulation in section
\ref{sec:CA}. Section \ref{sec:waves} presents the resulting wave patterns
and how they can be interpreted in the fundamental diagram. In the
thermodynamic limit, boundaries between the different wave regimes
are computed. Section \ref{sec-red} briefly addresses the impact of the
dawdling probability on the effective fundamental diagram.

Section \ref{sec:link} discusses a bottleneck simulation in the car-following
OV model, resulting in identical wave patterns as compared to the CA model.
In contrast to the CA model, analytical tools can be applied to explain the
wave selections. In particular, continuum theory is used to investigate
travelling waves in terms of plateau connections between equilibrium flow
solutions. A strong link between CA and OV models emerges.

Finally, some of the results are compared to work in the literature which
is related to bottleneck simulations (section \ref{sec:comp}), before we
draw some conclusions and mention planned future work in the last section.

\section{\label{sec:CA}Cellular Automata bottleneck simulation}
\begin{figure}[ht]
 \begin{center}
  \includegraphics[width=65mm]{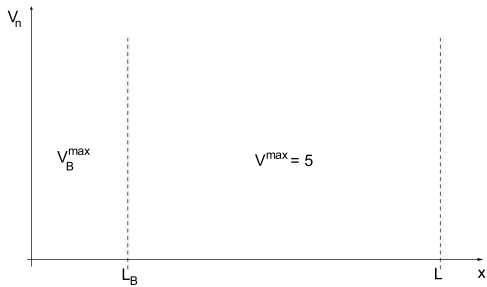}
\caption{The bottleneck simulation assumes periodic boundary conditions
(traffic on a loop). The ``open'' road ($L_B<x\leq L$) is modeled with the
conventional Nagel-Schreckenberg model \cite{nagel:92}. The bottleneck
($0\leq x\leq L_B$) uses the same model except for a smaller maximum speed
$v_B^{max}$. In the simulations, we set $L_B=200$ and $L=1000$.}
\label{fig1}
\end{center}
\end{figure}
For traffic on a loop (periodic boundary conditions) of length $L$, a
bottleneck of length $L_B$ is located at $0\leq x\leq L_B$ (Fig.\,\ref{fig1}).
The system is simulated with the Nagel-Schreckenberg (NS) CA model
\cite{nagel:92} for vanishing randomness ($p=0$) and a reduction in top speed
from $v^{max}=5$ on the ``open'' road to $v^{max}_B=3$
in the bottleneck. All other model parameters remain the same. Initially,
$N$ cars are randomly distributed along the road and the system is updated
up to $t=10^6$ time steps. We set $L_B=200$ and $L=1000$ and study the
emerging wave patterns. We also choose $p=0$ in order to avoid the jam
formation in the NS model, which would interfere with the stationary wave
patterns and complicate the analysis.

\begin{figure}[ht]
 \begin{center}
  \includegraphics[width=70mm]{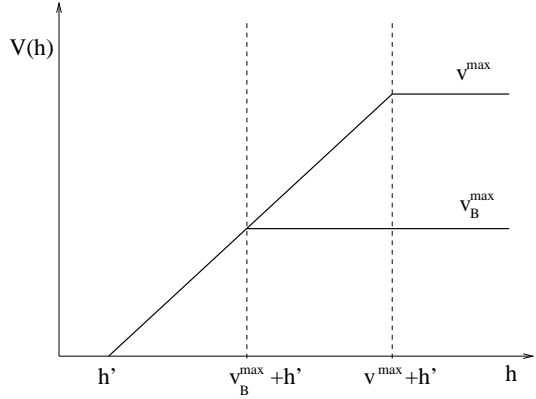}
\caption{The equilibrium velocity function of the CA model, velocity
$V$ as a function of the headway $h$ (distance) between the cars.
It is a piecewise linear function in the bottleneck and on the
open road with an upper bound of $v^{max}$ on the open road and
$v_{B}^{max}$ in the bottleneck. The functions are partially
identical which translates into partially identical fundamental
diagrams (see Fig.\,\ref{fig2}). Here, $h'$ represents the minimum
distance between the cars.}
\label{figvh}
\end{center}
\end{figure}

Figure \ref{figvh} is a qualitative plot of the equilibrium velocity function
used in the update rule of the NS CA model for both the bottleneck and the
open road. Here, $h'=1$, $v^{max}=5$ and $v^{max}_B=3$. Since the density
$\rho$ is continuous in this plot but the CA model is discrete in space,
$\rho$ must be interpreted in the thermodynamic limit as an average
density taken over many cells.

Note that the wave selection on a loop is fundamentally different from wave
selection on an open road with different conditions at the upstream and
downstream boundary, respectively \cite{kolo:98,chey:01,barlo:02}. On an
open road, travelling waves can only interact once with stationary structures
since they do not move around a loop. Therefore, an open road gives rise
to new wave solutions as compared to the loop, determined by the boundary
conditions \cite{kolo:98,chey:01,barlo:02}. This is a very different
setup from what is presented in this work.

\section{\label{sec:waves}Wave selection in the fundamental diagram}
\begin{figure}[ht]
 \begin{center}
  \includegraphics[width=80mm]{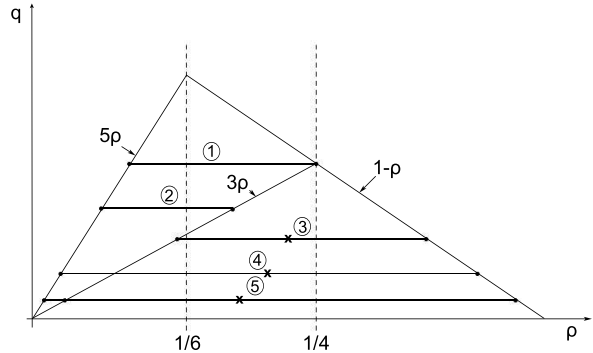}
\caption{Stationary shock waves visualized in the fundamental diagram:
From six possible wave connections (the trivial uniform flow solution
in Figs.\,\ref{fig6} and \ref{fig7} is not shown here), only three emerge
as dynamical solutions, displayed in Figs.\,\ref{fig4}-\ref{fig7}.}
\label{fig2}
\end{center}
\end{figure}

We will use the fundamental diagram (FD), i.e.\,flux versus density, and
kinematic wave theory to
interpret our numerical results. Based on the equilibrium velocity function
in Fig.\,\ref{figvh} and using the equilibrium flow-density relation
of uniform flow ($q$: flux, $\rho$: density)
\begin{equation}
q=v(\rho)\,\rho , \,\,\, \rho=1/h ,
\end{equation}
the FD is shown for both the bottleneck
\begin{equation}
q_B=\left\{
\begin{array}{lcl}
3\rho &;& 0\leq\rho\leq1/4 ,\\
1-\rho&;& 1/4<\rho\leq 1
\end{array}
\right.
\end{equation}
and the open road
\begin{equation}
q_o=\left\{
\begin{array}{lcl}
5\rho &;& 0\leq\rho\leq1/6 ,\\
1-\rho&;& 1/6<\rho\leq 1
\end{array}
\right.
\end{equation}
in Fig.\,\ref{fig2}.
The two curves merge into the same function for $\rho\geq 1/4$.

\begin{figure}[ht]
 \begin{center}
  \includegraphics[width=120mm]{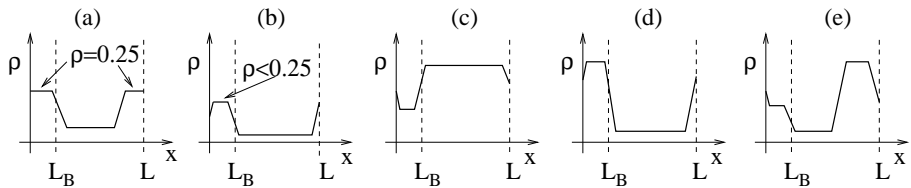}
\caption{Stationary wave patterns of Fig.\,\ref{fig2}. While
cases (a) and (b) occur in the simulations (cases (1) and (2) in
Fig.\,\ref{fig2}), as shown in
Figs.\,\ref{fig4} and \ref{fig5}, cases (c), (d), and (e) do not
emerge. This is a very close analogy to the wave pattern selection
of a bottleneck simulation with the optimal-velocity model
(see section \ref{sec:link}) \cite{ward:tgf05,wilson:05}.}
\label{fig3}
\end{center}
\end{figure}

Generally speaking, we could expect as many as six stationary wave solutions
for a bottleneck on a loop as $t\to\infty$. In the FD, five of them are
visualized as chords with zero gradient due to the requirement of vanishing
wave speed, based on kinematic wave theory \cite{berg:01}. By kinematic waves
we refer to density waves described by the first-order hyperbolic
Lighthill-Whitham model of traffic flow \cite{whitham}
\begin{equation}
\rho_t+q(\rho)_x=0 \Rightarrow \rho_t + q'(\rho)\rho_x =0 .
\end{equation}
For stationary waves, we have $\rho_t=0$ and the equation can be integrated
to yield
\begin{equation}
q(\rho(x))=const .
\label{qrho}
\end{equation}
Therefore, stationary shock waves, whose profiles are given by $\rho(x)$,
exhibit constant flux along the road, represented by a horizontal straight
line in the fundamenatal diagram. In Fig.\,\ref{fig2}, each chord is a
straight line between points on the fundamental diagrams, one point
representing the bottleneck and one representing the open-road FD.
They connect plateaus between the bottleneck and the open road (case 2, 3 and
4) in case of one stationary shock wave at each bottleneck boundary.
However, they
can also entail a plateau connection on the open road as in cases 1 and 5.
These five stationary wave patterns are shown in Fig.\,\ref{fig3} in terms of
density distribution along the loop.
In addition, there is the trivial wave solution of homogeneous uniform flow.
In principle, we could think of further wave patterns but we will restrict the
analysis to the simplest cases featured here.

\begin{figure}[ht]
 \begin{center}
  \includegraphics[width=80mm]{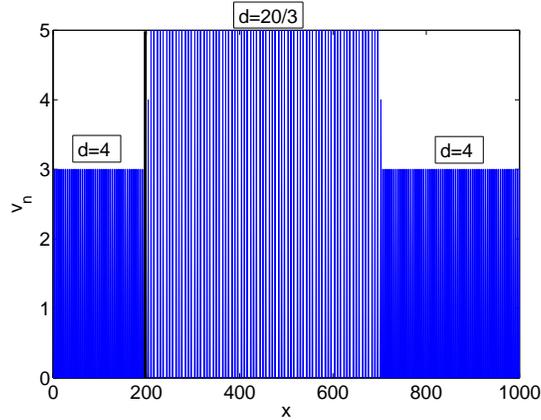}
\caption{Stationary wave pattern of bottleneck simulation with average
density $0.17<\rho<0.25$ (here: $\rho=0.20$): Two shocks emerge,
one at the downstream bottleneck boundary and one classical (Lax) shock
on the open road. The flow in the bottleneck is at its maximum.}
\label{fig4}
\end{center}
\end{figure}

\begin{figure}[ht]
 \begin{center}
  \includegraphics[width=80mm]{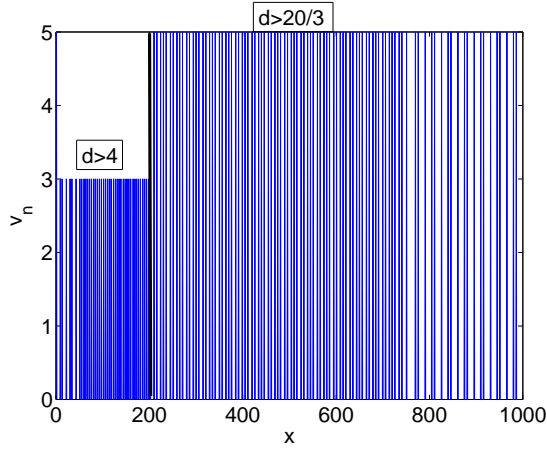}
\caption{Stationary wave pattern of bottleneck simulation with average
density $\rho<0.17$ (here: $\rho=0.142$): Two shocks emerge,
one at each bottleneck boundary.}
\label{fig5}
\end{center}
\end{figure}

\begin{figure}[ht]
 \begin{center}
  \includegraphics[width=80mm]{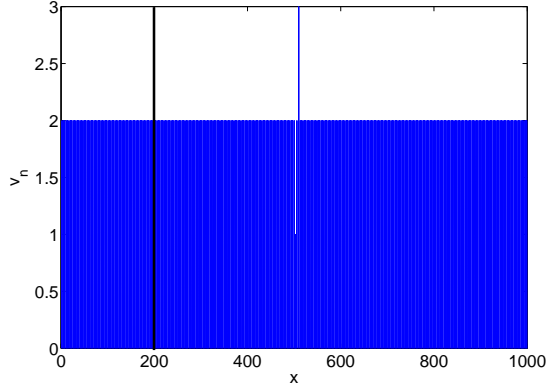}
\caption{Trivial flow solution for bottleneck simulation with average
density $\rho>0.25$ (here: $\rho=0.333\Rightarrow d\approx 3$).}
\label{fig6}
\end{center}
\end{figure}

\begin{figure}[ht]
 \begin{center}
  \includegraphics[width=80mm]{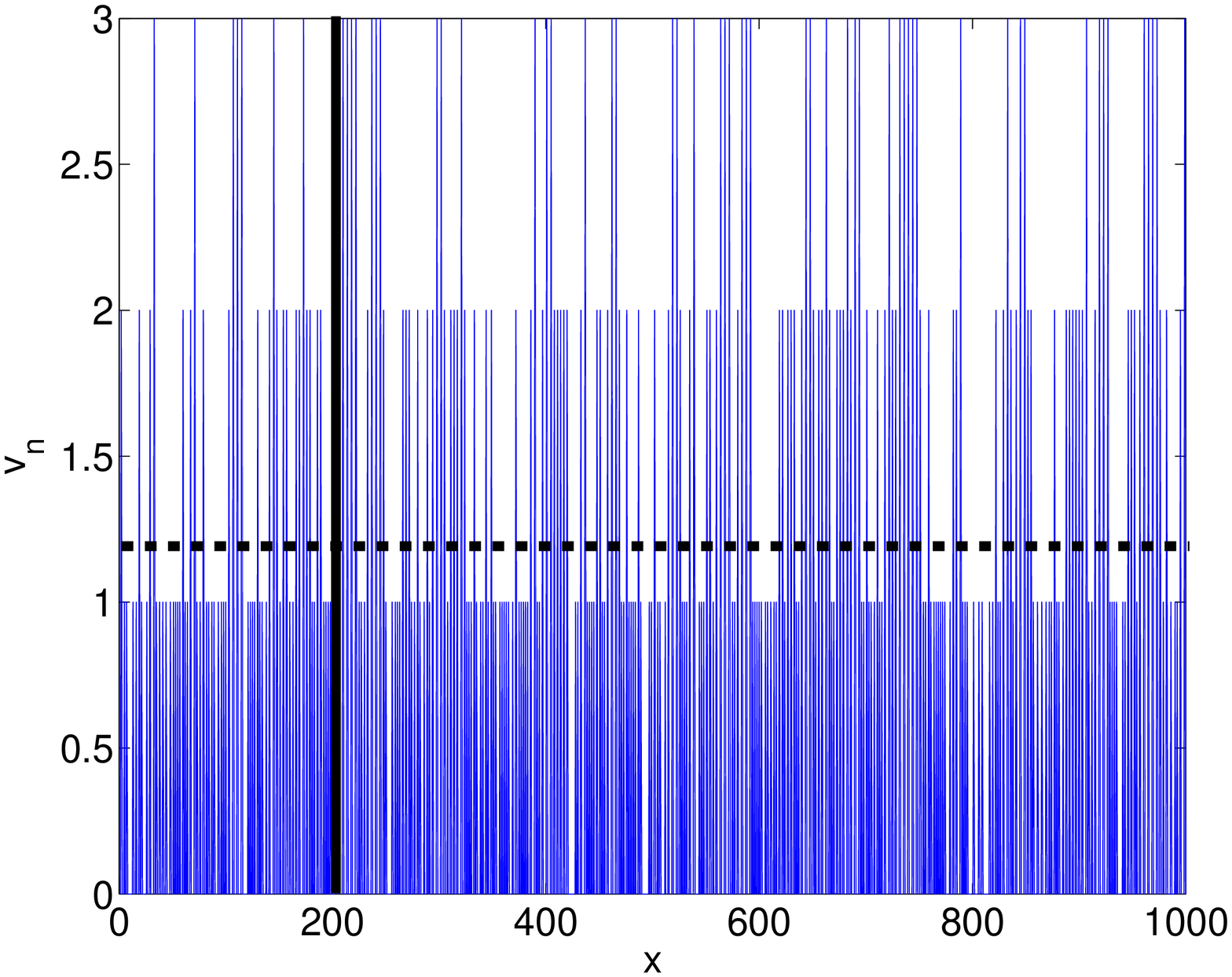}
\caption{Trivial flow solution for bottleneck simulation with average
density $\rho>0.25$ (here: $\rho=0.45\Rightarrow d\approx 2.22$).
The average speed $v\approx 1.22$ is visualized by the thick dotted line.
The average flux is exactly at $q=1-\rho=0.55$, as expected from a
trivial flow solution in the fundamental diagram. Even on the open-road
segment the speed does not exceed $v_B^{max}=3$. Therefore, the
bottleneck has no longer an impact on the flow solution.}
\label{fig7}
\end{center}
\end{figure}

We found in our simulations that only three wave patterns are selected from
this range of possible solutions. They consist of the following ($\rho$:
average density on the loop):
\begin{itemize}
\item Case 1: $0.17\leq \rho < 0.25$\\
Stationary wave pattern that connects two plateaus by one shock
at the downstream boundary of the bottleneck and one classical (Lax) shock
on the open road (see Figs.\,\ref{fig3}a and \ref{fig4}): The resulting
bottleneck headway (distance between cars) is exactly at $d_B=4$ and, therefore,
the bottleneck is at maximum flow. On the open
road we find the headway to be near $d_n=7$, or exactly $d_o=20/3$ on average.
\item Case 2: $0\leq \rho < 0.17$\\
Stationary wave pattern that connects two plateaus by a shock
at the upstream and downstream boundary of the bottleneck, respectively
(see Fig.\,\ref{fig3}b and \ref{fig5}): In the bottleneck $d_B>4$ and on the
open road $d_o>20/3$. It shall be stressed that it takes a very long time
for the system to reach
steady-state due to the small interaction of cars on the open road. Hence,
Fig.\,\ref{fig5} should be considered as a transient, quasi-steady state.
\item Case 3: $\rho > 0.25$\\
Trivial flow solution, i.e.\,homogeneous, uniform flow: Unless the average
headway $d$ is close to an integer number, as is the case in Fig.\,\ref{fig6},
the individual headways $d_n$ oscillate around the average headway $d$,
exhibited in Fig.\,\ref{fig7}. However, this is an effect solely due to the
discretization
of space, and the flow solution can still be considered uniform. In particular,
the average flow measured in Fig.\,\ref{fig7}, $q=0.55$, equals exactly the
uniform flow in the FD, as expected from an average density $\rho=0.45$, namely
$q=1-\rho=0.55$.
\end{itemize}

This leads to three open questions:
\begin{enumerate}
\item Why is there a transition between the structures at $\rho=0.17$
and $\rho=0.25$?
\item What determines the location of the classical shock on the open road
in case 1?
\item Why do we not observe the other wave patterns?
\end{enumerate}
We will now elaborate on all three questions.

The two headways in Fig.\,\ref{fig5} are determined by the conservation of cars
and by imposing zero wave speed (zero gradient of the chord in the FD),
described by Eq.\,(\ref{qrho}). This
can be written as two equations with two unknowns, the bottleneck headway $d_B$
and the open-road headway $d_o$. Neglecting finite size effects, conservation
of cars reads
\begin{equation}
\frac{L-L_B}{d_o}+\frac{L_B}{d_B}=N .
\label{eqn3}
\end{equation}
We find for the wave speed criterion (equal fluxes $q_B$ and $q_o$ in both
road segments)
\begin{equation}
q_B=q_o \Rightarrow 3\rho_B=5\rho_o \Rightarrow \frac{3}{d_B}=\frac{5}{d_o},
\label{eqn4}
\end{equation}
where $\rho_B$ and $\rho_o$ denote the bottleneck and open-road density,
respectively.

The system (\ref{eqn3})-(\ref{eqn4}) can be solved for $d_B$. It yields
\begin{equation}
d_B=\frac{\frac{3}{5}L+\frac{2}{5}L_B}{N}.
\end{equation}
In Fig.\,\ref{fig5}, we have $L=1000$, $L_B=200$, $N=142$ and, hence,
$d_B=4.79$. This equals $\rho_B=1/d_B=0.21$, which coincides with the
numerical value. The value for $d_o$ follows correspondingly.

The maximum amount of vehicles that the wave structure in Fig.\,\ref{fig5}
can support, however, is reached when $d_B=4.0$ and, determined by zero wave
speed, $d_o=20/3$. For $L_B=200$, this corresponds to an average density of
\begin{equation}
\rho=0.2\frac{1}{4}+0.8\frac{1}{20/3}=0.17.
\end{equation}
This is coincidentally close to $q=1/6$, the maximum of $q_o$, but varies
with the choice of $L_B$.
If we increase $\rho$ beyond this, the wave pattern in Fig.\,\ref{fig4} is
triggered, with a bottleneck headway exactly at $d_B=4$ and $d_o=20/3$ for the
other plateau value. This is shown by chord 1 in Fig.\,\ref{fig2}. The length
of the second plateau $L_p$ is now determined by the conservation of cars
alone. Setting $\rho=0.2$ in Fig.\,\ref{fig4}, we write
\begin{equation}
\rho= \frac{1}{4} (1-L_p/L) + \frac{1}{20/3} (L_p/L) \Rightarrow L_p=500,
\end{equation}
which is very close to the numerical value of $L\approx 510$. Note that finite
size effects impose limits on the accuracy of estimates. For $\rho \geq 0.25$,
which is the maximum of $q_B$,
the length $L_p$ of the open-road plateau equals zero
and this wave pattern must vanish. We are left with the trivial flow solution
of Figs.\,\ref{fig6} and \ref{fig7} since the average density exceeds the
maximum density that can support the wave structure in Fig.\,\ref{fig4}.

Thus, we have answered questions 1 and 2. Before we turn to question 3
in section \ref{sec:link}, let us study the effective fundamental diagrams
obtained in the simulations. Here, we will only briefly discuss the impact
of $p>0$ on the fundamental diagram and leave a more detailed study to
future research.

\section{\label{sec-red}Impact of the dawdling probability on the \\
fundamental diagram}
\begin{figure}[ht]
 \begin{center}
  \includegraphics[width=80mm]{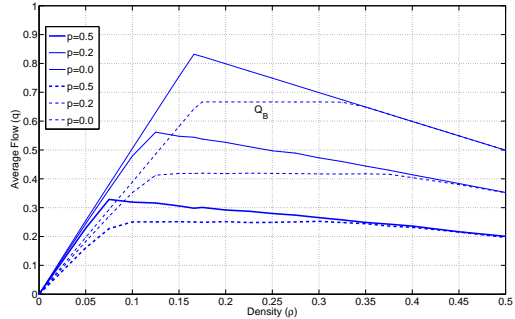}
\caption{Effective fundamental diagrams (dashed lines) for three different
values of the dawdling probability $p$ from a simulation with $L=1000$,
$L_B=200$, $v^{max}=5$ and $v_{B}^{max}=2$. Three distinct regimes can be
identified. For comparison, the FD for the same simulations without a
bottleneck (solid line; $L_B=0$) exhibit only two distinct regimes.}
\label{fig1p}
\end{center}
\end{figure}

The well-defined wave patterns for $p=0$ overlap with instabilities of the
flow when $p>0$. Therefore, the dawdling probability $p$ has a significant
impact on the effective fundamental diagram, which is defined as the average
flux on the bottleneck loop versus the average density, as shown in
Fig.\,\ref{fig1p}. It shows three separate values of $p$ ($p=0.0;0.2;0.5$).
To plot the figure, the parameters were chosen as $L=1000$, $L_B=200$,
$v^{max}=5$ and $v_B^{max}=2$. As the dawdling probability is increased,
there is a decline in the flux for both parts of the road. This is expected
since the vehicles are more likely to reduce their speed randomly as $p$
increases. The flow also becomes more unstable.

When $p=0$, it can be shown analytically that the effective FD is piecewise
linear and consists of three segments, joined at the critical values
$\rho_{c1}=13/75 \approx 0.1733$ and $\rho_{c2}=1/3$. These three linear
regimes correspond to the three wave patterns found in the previous
section. The critical values for $\rho$ are determined in a fashion similar
to the critical density values in section \ref{sec:waves}. The linear
function
\begin{equation}
Q(\rho)=\frac{50}{13}\rho
\end{equation}
for $0\leq \rho\leq 13/75$ is also derived from conservation of cars
and zero wave speed. The intermediate constant function, determined
by the maximum of $q_B$ for case 1 in Fig.\,\ref{fig2}, then joins
onto the FD of the open road (no bottleneck; $L_B=0$).
Hence, we have a complete understanding of the
effective FD for a bottleneck with $p=0$.

However, for $p>0$ we resort to numerical results only which again exhibit
three distinct regimes, or phases, with rather blurry transitions between them.
The occurrence of such three phases has
been discovered in previous CA simulations and will we address some of these
in section \ref{sec:comp}.

\section{\label{sec:link}Link to optimal-velocity models}
While the above first-order analysis explains the transition between the three
wave patterns which are observed in the numerical simulations, it does not
answer why they are selected or why the remaining wave patterns in
Fig.\,\ref{fig2}
do not occur. In this section, we will draw a parallel to the microscopic
car-following optimal-velocity (OV) model. We will simulate a bottleneck in
the OV model in a way similar to the CA model and analyze how it compares
to the CA results. It turns out that the same three wave patterns appear again
and occur exclusively. Using the corresponding continuum model, we can explain
this phenomenon with second-order analytical methods.

\subsection{Bottleneck simulation in the OV model}
In the optimal-velocity model \cite{bando:95}, the acceleration of each
vehicle is determined by an equation of motion that includes an optimal
(or desired) velocity $V$ which depends on the distance, $h_n=\Delta x_n$,
to the preceding vehicle.

The governing equation reads
\begin{equation}
\ddot{x}_n = a[V(\Delta x_n) - \dot{x}_n]
\label{eqOV1}
\end{equation}
where $\Delta x_n = x_{n+1} - x_n$, and $n$ denotes the index of each
vehicle at position $x_n$.  The velocity is represented by $\dot{x}_n$ and
acceleration by $\ddot{x}_n$.

In order to be consistent with the CA model, we use a piecewise linear
optimal-velocity function with a maximum velocity in the bottleneck at $60\%$
of the maximum velocity on the open road, again shown qualitatively in
Fig.\,\ref{figvh}. For simplicity, we used $v_{B}^{max} = 0.6$ and
$v^{max} = 1.0$ with an offset $h' = 0.1$, representing the distance a
vehicle occupies in a jam. Note that we are working with a dimensionless
model just as it was the case for the CA model.

This results in an optimal-velocity function (bottleneck: subscript {\em B};
open road: subscript {\em o})
\begin{eqnarray}
V_B(h)=\left\{
\begin{array}{lcl}
h-0.1 &;& h<0.7 ,\\
0.6 &;& h\geq0.7
\end{array}
\right.\\
V_o(h)=\left\{
\begin{array}{lcl}
h-0.1 &;& h<1.1 ,\\
1.0 &;& h\geq1.1
\end{array}
\right.
\end{eqnarray}
and the fluxes
\begin{eqnarray}
q_B=\left\{
\begin{array}{lcl}
1-0.1\rho &;& \rho> 1/0.7 ,\\
0.6\rho &;& \rho\leq 1/0.7
\label{ovsimq1}
\end{array}
\right.\\
q_o=\left\{
\begin{array}{lcl}
1-0.1\rho &;& \rho> 1/1.1 ,\\
\rho &;& \rho\leq 1/1.1 .
\label{ovsimq2}
\end{array}
\right.
\end{eqnarray}
The sensitivity $a$ is chosen so that we guarantee linear stability. The
flow stability is determined by \cite{bando:95}
\begin{equation}
V'(h) < a/2 ,
\end{equation}
which also defines a range of headways for which the model is unstable.
For the simulations performed, $a=2.0$ was chosen to maintain stable flow.

Initially, we place $N$ vehicles randomly along a loop of length $L$ with a
bottleneck of length $L_B$ and let the system evolve.

\subsection{Wave selection in the fundamental diagram}
Using the above equations for the flux in the bottleneck and on the open road,
the fundamental diagram is (qualitatively) identical to that of the CA model
in Fig. \ref{fig2}. After a variety of simulations are carried out, the same
three wave patterns which occurred in the CA model, also emerge in the OV model,
while other wave patterns never appear. These three wave patterns also form
three traffic phases which are qualitatively identical to those of the CA
model (see Fig.\,\ref{fig1p}).

We find: i) Two shocks at low density, one occurring at either end of the
bottleneck, corresponding to case 2 in the CA model; ii) two shocks at
medium density, one occurring at the upstream bottleneck boundary and one
(classical Lax shock) appearing on the open road (case 1 in the CA model);
iii) uniform, homogeneous flow (case 3 in the CA model).

These numerical results, including the critical densities of each regime, can
again be discussed with first-order analytical methods, namely conservation of
cars and zero wave speed, i.e.\,kinematic wave theory. This analysis explains
again the connection between the wave patterns observed in the (OV) model but
it does not explain why these are the only wave connections
selected in Fig.\,\ref{fig2}. In order to investigate this phenomenon in more
detail, we will now use continuum theory and analyse the travelling-wave phase
plane of the continuum analogue of the OV model.

\subsection{Travelling-wave phase-plane analysis}
Previous work in continuum theory by Berg \textit{et al.} \cite{berg:00} and
Lee \textit{et al.} \cite{lee:01} will now be used to analyse the
travelling-wave patterns expected from the fundamental diagram of the OV
model and, hence, the CA model. A priori, it is unclear whether wave patterns
may occur in the OV model which cannot be found in the CA model. Hence,
second-order continuum theory will be applied to analyse {\em all} plateau
connections in the OV model.

Lee \textit{et al.} \cite{lee:01} derived a second-order continuum model of
the OV model, using density $\rho$ and velocity $v$ as the two variables.
Using the conservation of cars
\begin{equation}
\rho_t + (v\rho)_x = 0
\label{conseq}
\end{equation}
and the dynamic equation involving the optimal velocity
\begin{equation}
v_t + vv_x = a[\hat V(\rho) - v] + \frac{a\hat V'(\rho)}{2\rho}\rho_x
+ \frac{a}{6\rho^2}v_{xx} ,
\label{dyneq}
\end{equation}
the stability of the shock tails will now be analysed, meaning the stability
of the equilibrium points (EP) in the corresponding travelling-wave phase
plane. Here, we define $\hat V(\rho)=V(1/ \rho)=V(h)$.

\subsubsection{Phase-plane analysis under constant optimal velocity}
A constant optimal velocity, $\hat V'(\rho)=0$, refers to the positive slopes
of the fundamental diagram where the flux is found from $q=v^{max}\rho$ in
Eqs.\,(\ref{ovsimq1})--(\ref{ovsimq2}). Therefore, we have
$V(h)=V(1/\rho)=\hat V(\rho)=v^{max}$ and $\hat V'(\rho)=0$, where $v^{max}$ is
the constant maximum optimal velocity. Note that the same analysis is applied
to the bottleneck regime by substituting $v_B^{max}$ for $v^{max}$.

With these parameters and assuming stationary shock waves ($v_t = \rho_t = 0$),
Eq.\,(\ref{dyneq}) becomes
\begin{equation}
vv_x = a[v^{max} - v] + \frac{a}{6\rho^2}v_{xx}.
\end{equation}
Using the definition for the flux, $q = v\rho$, and linearization via
$v=v^{max}+\hat{v}$ yields
\begin{equation}
v^{max}\hat{v}_x = -a\hat{v} + \frac{a {v^{max}}^2}{6q^2}\hat{v}_{xx}.
\end{equation}
Here, the flux $q$ is given by the boundary conditions at infinity.

This equation can now be written as two first-order ordinary differential
equations
\begin{eqnarray}
v_x &=& w , \\
w_x &=& \frac{6q^2}{a {v^{max}}^2}[v^{max}w + av] ,
\end{eqnarray}
where we have dropped the ``hat'' notation for $v$.

To find the stability of the equilibrium points, the eigenvalues of the
preceding equations are found by rewriting the equations in matrix form
\begin{equation}
\left(
\begin{array}{c}
v_x \\
w_x
\end{array}
\right)
=
\left(
\begin{array}{cc}
0 & 1 \\
\frac{6q^2}{{v^{max}}^2} & \frac{6q^2}{a v^{max}}
\end{array}
\right)
\left(
\begin{array}{c}
v \\
w
\end{array}
\right) .
\end{equation}
Trying the ansatz $v,w\sim \exp(\lambda x)$,
this yields the following two roots of the characteristic equation
\begin{equation}
\lambda_{1,2} = \frac{3q^2}{a v^{max}} \pm \sqrt{\left( \frac{3q^2}{a v^{max}}
\right)^2 + \frac{6q^2}{{v^{max}}^2}} .
\end{equation}
Since $q,v^{max}>0$ holds, the root is always larger than the term preceding the
root and the eigenvalues represent a saddle in the phase plane of $v$ and
$v_{x}$. This allows wave connections both i) from an upstream plateau
(equilibrium point) to this equilibrium point, representing the second plateau,
along the stable manifold and ii) from this plateau to another plateau along the
unstable manifold. Hence, wave connections are possible along the positive
slopes of the fundamental diagram, as in case 2 in Fig.\,\ref{fig2}, which can
now be interpreted as a saddle-saddle connection.

\subsubsection{Phase-plane analysis for an optimal velocity dependent upon
headway}
An analysis is now performed for the monotonically decreasing part
of the fundamental diagram where the OV function does vary with headway
and, consequently, with density. Here, we have $V(h)=\hat V(\rho)= 1/\rho-0.1$.
The same analysis is carried out as before, however, this time for the
dynamical equation
\begin{equation}
vv_x = a\left[\frac{1}{\rho} - 0.1 - v\right] - \frac{a}{2\rho^3}\rho_x
+ \frac{a}{6\rho^2}v_{xx} .
\end{equation}
Using $q = v\rho$ again, and hence $\rho_x = \frac{-qv_x}{v^2}$, we obtain
\begin{equation}
v v_x = a\left[ \left( \frac{1}{q} - 1 \right)v - 0.1 \right]
+ \frac{av}{2q^2}v_x + \frac{av^2}{6q^2}v_{xx} .
\end{equation}
Now let $v = \bar{v} + \hat{v}$ and $\bar{v} = \frac{0.1}{1/q-1}$, we find
\begin{equation}
\bar{v}\hat{v}_x = a\left( \frac{1}{q} - 1 \right)\hat{v}
+ \frac{a\bar{v}}{2q^2}\hat{v}_x + \frac{a\bar{v}^2}{6q^2}\hat{v}_{xx} .
\end{equation}
Writing this again in terms of two first-order differential equations, we obtain
\begin{eqnarray}
v_x &=& w, \\
w_x &=& \frac{6q^2}{a\bar{v}^2} \left[\bar{v}\left(1 - \frac{a}{2q^2}\right)w
- a\left(\frac{1}{q} - 1\right)\hat{v} \right] .
\end{eqnarray}
To find the stability of the equilibrium points, the eigenvalues of the
preceding equations are determined via matrix notation
\begin{equation}
\left(
\begin{array}{c}
v_x \\
w_x
\end{array}
\right)
=
\left(
\begin{array}{cc}
0 & 1 \\
\frac{6q^2}{\bar{v}^2}\left(1 - \frac{1}{q}\right) \,\,\,\,\,\,\,&
\frac{6q^2}{a\bar{v}}\left(1 - \frac{a}{2q^2}\right)
\end{array}
\right)
\left(
\begin{array}{c}
v \\
w
\end{array}
\right)
\end{equation}
and so
\begin{eqnarray}
\lambda_{1,2} &=& \frac{3q^2}{a\bar{v}}\left(1 - \frac{a}{2q^2}\right)
\nonumber \\
              &\pm& \sqrt{\left[ \frac{3q^2}{a\bar{v}}\left(1 -
\frac{a}{2q^2}\right) \right]^2 + \frac{6q^2}{\bar{v}^2}\left(1 -
\frac{1}{q}\right)} .
\label{lambda2}
\end{eqnarray}
Since $0 < q < 1$ holds,
it can be seen that $\left(1 - \frac{1}{q}\right) < 0$ and the
argument under the square root can be positive or negative.  In any case,
stability is governed by the sign of the real part of
$\lambda_{1,2}$ and that is always determined by
$\frac{3q^2}{a\bar{v}}\left(1-\frac{a}{2q^2}\right)$ since the root is either
purely imaginary or real but smaller than
$\frac{3q^2}{a\bar{v}}\left(1-\frac{a}{2q^2}\right)$. Since $q,a,\bar v>0$
holds, the real part will either be greater than or less than zero depending
on the term $\left(1 - \frac{a}{2q^2}\right)$.  If $a \geq 2$ and $0 < q < 1$,
which is the case in our simulations (see Eqs.\,(\ref{ovsimq1}) and
(\ref{ovsimq2}) where $q\leq 1/1.1$), the real part will be less than zero.
Hence, the equilibrium point
(EP) is a stable node (or spiral). Therefore, we can connect from an upstream
plateau to this EP but we cannot connect from this EP to another equilibrium
point downstream to form a wave pattern. This prevents the formation of
patterns 3, 4 and 5 in Fig.\,\ref{fig2}, which would involve a non-permissible
connection starting from an EP of the monotonically decreasing branch of the
FD.

For the same reasons, case 1 in Fig.\,\ref{fig2} must be considered as a
limiting scenario of the above-mentioned saddle-saddle connection of case 2
between the monotonically increasing branch of the bottleneck FD and the
open-road FD, respectively.

Future work will investigate a bottleneck simulation for unstable flow where
$a<2$ holds. Then it is possible to find unstable nodes in Eq.\,(\ref{lambda2})
since we can have
\begin{equation}
1-\frac{a}{2q^2}>1
\end{equation}
or, equivalently,
\begin{equation}
a<2q^2 .
\end{equation}

\subsubsection{Further thoughts}
At this stage, it is unclear how the instability of the flow will affect the
wave patterns and whether some stable wave patterns will persist.

Also, it should be noted that a similar bottleneck simulation with the
(linearly
stable) OV model \cite{bando:95} has been carried out by Wilson {\em et al.}
\cite{ward:tgf05,wilson:05}. Here, the bottleneck was simulated by a reduction
factor in the optimal-velocity function, resulting in new wave patterns.
A coarse-graining method reveals that second-order theory might be insufficient
to explain all wave patterns at the microscopic level. However, a higher-order
continuum model is not available. Moreover, if one follows the idea of
Berg {\em et al.} \cite{berg:00} and extended their continuum model to
higher order, it would involve second-order (and higher) derivatives of
$\hat V(\rho)$ and they all vanish for a piecewise linear OV function so that
a higher-order modelling approach would fail in the context of this paper.

Furthermore, the continuum model of Berg {\em et al.} completely fails in
explaining the saddle-saddle connections since one ends up with an insufficient
first-order model
\begin{equation}
vv_x = a[v^{max} - v]
\end{equation}
along those branches of the FD. This clearly shows the advantage of the
Lee {\em et al.} model, adding to its advantage over the Berg {\em et al.}
model in terms of matching the linear stability criterion of the discrete
OV model at short wave lengths.

In summary, kinematic wave theory alone cannot fully explain the selection of
stationary shocks in the bottleneck simulation. Surely, stationary
wave patterns can be expected at the boundaries due to a non-smooth change in
model parameters. However, it is the dynamics of the model which determine
the actual observable patterns, and these resemble each other in the OV and
CA model simulations. This includes the appearance of three traffic phases in
either model.

\section{\label{sec:comp}Comparison to open-system and other \\
non-homogeneous CA models}
There have been several publications about CA models in the literature which
contain either two \cite{chey:01b}, three
\cite{yukawa:94,chey:01,barlo:02,kolo:98,lakatos:05} or several traffic phases
\cite{popkov:99}, as defined in section \ref{sec-red}.

Yukawa {\em et al.} \cite{yukawa:94} simulated a blockage on a loop in the
rule-184 CA model, having a maximum speed $v^{max}=1$. At exactly one site
along the loop, the hopping probability of an otherwise deterministic model
is reduced from $r=1$ to $0<r<1$. The authors find three traffic phases,
depending on the average traffic density and the hopping parameter $r$.
In order of increasing density, they are: i) free, ii) constant-flow and
iii) jam phase. This resembles the three wave patterns in this work
(see section \ref{sec-red}), where
the constant-flow phase is determined by $q_B^{max}=3/4$ in the regime
$0.17\leq \rho \leq 0.25$. However, while the free phase in the rule-184 CA
model is the same with or without a blockage site, this is not the case
in our simulations. In addition, the present work considers a fully
deterministic CA model and focuses on its emerging wave patterns. Moreover,
the bottleneck simulations in section \ref{sec:waves} will change as
$L_B \to 1$ due to finite size effects since the formation of plateaus
within the bottleneck will disappear. Therefore, planned future work will
investigate how the limit $L_B\to 1$ will affect the results both for
$v_B^{max}=3$ and $v_B^{max}=v^{max}=5$, and a non-zero dawdling probability
$p>0$ in the bottleneck only.

Lakatos {\em et al.} \cite{lakatos:05} extended the idea of Yukawa {\em et al.}
to multiple blockage sites in a totally asymmetric simple exclusion process
(TASEP) model on an open road. Approximate mean-field theory provides good
agreement between the numerical results and Monte Carlo simulations.
Phase diagrams are derived, which contain again three phases, but wave
patterns are not discussed due to the lack thereof. More importantly, the
boundary conditions mainly drive the dynamics of this open system, which
brings us to the next set of simulations.

The stationary wave patterns of the present bottleneck simulation would
change dramatically if the loop was replaced by an open road with fixed but
different up- and downstream boundary conditions
\cite{popkov:99,kolo:98,chey:01,chey:01b,barlo:02}.
The constraint of conservation of cars, which enables the analytical
computations of phase boundaries between the three wave patterns (see also
\cite{yukawa:94}), disappears and the analysis of the model becomes more
challenging. Depending on the model, three \cite{kolo:98} or more phases
\cite{popkov:99} can be found. Here, stationary wave patterns are determined
by the boundary conditions alone and can even be triggered in an otherwise
stable model, with or without a bottleneck.
In contrast, such patterns may or may not exist on the loop,
depending on the stability of the model. This fundamental difference is
exhibited for the OV model on a loop and open road by the work of Bando
{\em et al.} \cite{bando:95} and Berg {\em et al.} \cite{berg:01b},
respectively. Therefore, future work will also investigate the bottleneck
setup of the present paper under open-system boundary conditions similar to
\cite{popkov:99,kolo:98,chey:01,chey:01b,barlo:02}.

\section{\label{sec:conclusion}Conclusion and future work}
In this paper, the dynamics of a bottleneck simulation exhibit a link between
cellular automata and optimal-velocity traffic models. Three wave patterns,
which correspond to three distinct traffic phases in the fundamental diagram,
occur in either model and are qualitatively identical. This is surprising in
some sense since it connects a model, which is discrete in space and time, to
a model, which is continuous in space and time. The emerging wave patterns
were analysed and explained with second-order continuum theory.

Future work will focus on a CA model with non-zero dawdling probabilities
$0<p<1$. However,
this entails the formation of jams, which overlap with the stationary wave
patterns. On the other side, similarities to the work of Yukawa {\em et al.}
\cite{yukawa:94} can be expected to emerge from these simulations.

Moreover, the stationary wave patterns which occur on a long straight road,
containing a bottleneck and different up- and downstream boundary conditions,
will be studied. Here,
travelling waves cannot interact multiple times with stationary patterns.
This should give rise to new wave solutions as compared to the loop,
determined by the boundary conditions. This resembles research presented in
\cite{kolo:98,chey:01,barlo:02} but is fundamentally different from what
was presented in this work.

\section*{Acknowledgement}
Peter Berg is supported by an NSERC Discovery Grant. We would like to
thank Y.~Sugiyama for bringing bottleneck simulations to the authors
attention, B.S.~Kerner and R.E.~Wilson for the valuable discussions.

\end{document}